\documentclass[14pt]{revtex4}

\usepackage{graphicx,epstopdf}

\begin{document}

\title{An attempt for an Emergent Scenario with Modified Chaplygin Gas}

\author{Sourav Dutta$^1$\footnote {sduttaju@gmail.com}}
\author{Sudeshna Mukerji$^2$\footnote {mukerjisudeshna@gmail.com}}
\author{Subenoy Chakraborty$^1$\footnote {schakraborty@math.jdvu.ac.in}}
\affiliation{$^1$Department of Mathematics, Jadavpur University, Kolkata-700032, West Bengal, India\\}


\begin{abstract}
The present work is an attempt for emergent universe scenario with modified Chaplygin gas. The universe is chosen as spatially flat FRW space-time with modified Chaplygin gas as the only cosmic substratum. It is found that emergent scenario is possible for some specific (unrealistic) choice of the parameters in the equation of state for modified Chaplygin gas.
\end{abstract}
\maketitle

The inability of Einstein's general theory of relativity at zero volume leads to the well known big bang singularity in standard cosmology. To overrule this initial discomfortable   situation various cosmological scenarios have been proposed and are classified as bouncing universes or the emergent universes. Here we shall choose the second option which results from searching for singularity free inflationary scenario  in the background of classical general relativity. In a word, emergent universe is a model universe, ever existing with almost static behavior in the infinite past $(t\rightarrow -\infty)$ (gradually evolves into inflationary stage) and having no time--like singularity. Also the modern and extended version of the original Lemaitre--Eddington universe can be identified as the emergent universe scenario.\\\\
  Long back in 1967, Harrison [1] showed a model of the closed universe containing radiation, which approaches the state of an Einstein static model asymptotically (i.e, $t\rightarrow -\infty)$. This kind of model was again re investigated, after a long gap by Ellis and collaborators [2, 3]. Although they were not able to obtain exact solutions but presented closed universes with a minimally coupled scalar field $\phi$ having typical self interacting potential and possibly some ordinary matter with equation of state $p=w\rho$, ($-\frac{1}{3}\leq w\leq 1$), whose behavior alike that of an emergent universe was highlighted. Then in starobinsky model Mukherjee et al. [4] derived solutions for flat FRW space--time having emergent character in infinite past. Subsequently Mukherjee and associates [5] presented a general framework for an emergent universe model with an adhoc equation of state connecting the pressure and density, having exotic nature in some cases. These models are interesting as they can be cited as specific examples of nonsingular (i.e, geometrically complete) inflationary universes. Also it is worthy to mention here that entropy considerations favour the Einstein static model as the initial state for our universe [6, 7]. There after, a series of works [8--15] have been done to formulate emergent universe in different gravity models and also for various types of matter. Very recently, emergent scenario has been formulated with some interesting physical aspects: The idea of quantum tunneling [16] has been used for the decay of a scalar field having initial static state as false vacuum, to a state of true vacuum. Secondly, a model of an emergent universe has been formulated in the background of non--equilibrium thermodynamical prescription with dissipation due to particle creation mechanism [17]. Very recently, Paul et al. [18] has formulated emergent universe with interacting fields. Finally, Pavon  et al. [19, 20] has studied the emergent scenario from thermodynamical view point. They have examined the validity of the generalized second law of thermodynamics during the transition from a generic initial Einstein static phase to the inflationary phase and also during the transition from inflationary era to the standard radiation dominated era.\\\\
		
		Mixed exotic fluid known as modified Chaplygin gas [20] has the equation of state [21, 22]
		
		\begin{equation}
		p=A\rho-\frac{B}{\rho^n}~,~~~~~~~~~~~~~~~~~~~~~~~~~~0<n\leq 1.
		\end{equation}
		
		This equation of state shows barotropic perfect fluid $p=A\rho$, at very early phase (when the scale factor  a(t) is vanishingly small) while it approaches $\Lambda CDM$ model when the scale factor is infinitely large. It shows a mixture at all stages. Note that at some intermediate stage the pressure vanishes and the matter content is equivalent to pure dust. Further, this typical model is equivalent to a self interacting scalar field from field theoretic point of view. In the present paper we shall examine whether emergent scenario is possible for FRW model of the universe with matter content as modified Chaplygin gas (MCG).\\\\
		
		For homogeneous and isotropic flat FRW model of the universe, the Einstein field equations are (choosing $8\pi G=1$) \\
		\begin{equation}
		3H^2=\rho~,~~~~~~~~~~~~~~~2\dot{H}=-(\rho+p),
		\end{equation}
		with energy conservation relation:\\
		\begin{equation}
		\dot{\rho}+3H(\rho+p)=0.
		\end{equation}
		Using (1) in(3) one can integrate $\rho$ as
		\begin{equation}
		\rho=\left[\frac{B}{1+A}+\frac{c}{a^{3\mu}}\right]^{\frac{1}{n+1}},
		\end{equation}
		with $c>0$, a constant of integration.\\
		
		Now using this $\rho$ in the first Friedmann equation in (2) one can integrate to obtain cosmic time as a function of the scale factor as
		\begin{equation}
		\frac{\sqrt{3}}{2}(1+A)c^{\alpha}(t-t_0)=a^{\frac{3(1+A)}{2}} ~_2F_1\left[\alpha, \alpha, 1+\alpha, -\frac{B}{C(1+A)} a^{\frac{3(1+A)}{2 \alpha}}\right],
		\end{equation}
  where $\alpha =\frac{1}{2(1+n)}$ and $_2F_1$ is the usual hypergeometric function.\\

We shall now analyze the two asymptotic cases:\\

{\bf I. \underline{When the scale factor $'a'$ is very small}: }\\

For small $'a '$, $\rho$ can be approximated from equation(4) and $p$ from equation (1) as

\begin{equation}
\rho \cong \left(\frac{\rho_0}{A+1}\right)^{\frac{1}{n+1}} a^{-3(A+1)}+\frac{B}{(n+1)(A+1)^{\frac{1}{n+1}}\rho_0^{\left(\frac{n}{n+1}\right)}}a^{3(1+A)n} \equiv \rho_{1i}+\rho_{2i}.
\end{equation}\\

\begin{eqnarray}
p &\cong& \frac{A \rho_0^{\frac{1}{n+1}}}{(A+1)^{\frac{1}{n+1}}a^{3(A+1)}}-\frac{B\left[1+n(A+1)\right]}{(n+1) (A+1)^{\frac{1}{n+1}} \rho_0^{\frac{n}{n+1}}} a^{3n(1+A)},\nonumber\\\\
&\equiv&  p_{1i}+p_{2i}.\nonumber
\end{eqnarray}\\

{\bf II. \underline{When the scale factor $'a'$ has infinitely large value}: }\\

Similarly for large $'a'$, $\rho$ and $p$ are approximated from equation (4) and (1) respectively as

\begin{eqnarray}
\rho &\cong& \left(\frac{B}{A+1}\right)^{\frac{1}{n+1}} + \frac{\rho_0}{(n+1)B} \left(\frac{B}{A+1}\right)^{\frac{1}{n+1}} a^{-3 \mu},\nonumber\\\\
&\equiv& \rho_{1f}+\rho_{2f}.\nonumber
\end{eqnarray}\\

\begin{eqnarray}
p &\cong& -\frac{1}{(A+1)^{\frac{1}{n+1}}} + \frac{n + (n+1)A}{(A+1)^{\frac{1}{n+1}}} \frac{\rho_0}{(n+1)B} a^{-3 \mu},\nonumber\\\\
 &\equiv& p_{1f}+p_{2f}.\nonumber
\end{eqnarray}\\

Thus in the asymptotic limits the components of energy density and pressure can be expressed as sum of two non--interacting barotropic fluids having equation of states:\\
\begin{equation}
w_{1i}=A~,~~~~~~~~w_{2i}=- \left[1+n(A+1)\right],
\end{equation}\\
\begin{equation}
w_{1f}= -B^{-\left(\frac{1}{n+1}\right)}~,~~~~~~~~~w_{2f}= \frac{n+ (n+1)A}{B^{\left(\frac{1}{n+1}\right)}}.
\end{equation}\\

Thus MCG can be considered in the asymptotic limits as two barotropic fluids of constant equation of state of which one is exotic in nature. However, one can consider the two fluids in question (in the asymptotic limit) may be interacting with separate equation of state as \\

\begin{eqnarray}
\dot{\rho_{1i}}+ 3(\rho_{1i}+p_{1i})H&=& Q_i,\nonumber\\\\
\dot{\rho_{2i}}+ 3(\rho_{2i}+p_{2i})H&= & -Q_i,\nonumber
\end{eqnarray}

and

\begin{eqnarray}
\dot{\rho_{1f}}+ 3(\rho_{1f}+p_{1f})H&=& Q_f,\nonumber\\\\
\dot{\rho_{2f}}+ 3(\rho_{2f}+p_{2f})H&= & -Q_f,\nonumber
\end{eqnarray}
where $Q_i$ and $Q_f$ represent the interaction term.\\

$Q_i>0$ indicates a flow of energy from fluid 2 (having energy density $\rho_{2i}$) to fluid 1 (having energy density $\rho_{1i}$) and similarly for $Q_f$ also. Further, one can rewrite the conservation equations (12) and (13) as

\begin{eqnarray}
\dot{\rho_{1i}}+ 3H(1+w^{eff}_{1i})\rho_{1i}&=& 0,\nonumber\\\\
\dot{\rho_{2i}}+3H(1+w^{eff}_{2i})\rho_{2i}&=& 0 ,\nonumber
\end{eqnarray}

and

\begin{eqnarray}
\dot{\rho_{1f}}+ 3H(1+w^{eff}_{1f})\rho_{1f}&=& 0,\nonumber\\\\
\dot{\rho_{2f}}+3H(1+w^{eff}_{2f})\rho_{2f}&=& 0 ,\nonumber
\end{eqnarray}

with

\begin{eqnarray}
w^{eff}_{1i}&=&w_{1i}-\frac{Q_i}{3H\rho_{1i}}~~,~w^{eff}_{2i}=w_{2i}+\frac{Q_i}{3H\rho_{2i}},\nonumber\\\\
w^{eff}_{1f}&=&w_{1f}-\frac{Q_f}{3H\rho_{1f}}~~,~w^{eff}_{2f}=w_{2f}+\frac{Q_f}{3H\rho_{2f}}.\nonumber
\end{eqnarray}

The above conservation equations show that the fluids may be considered as non--interacting at the cost of variable equation of state.\\

One should note that in integrating equation (3) to have equation (4) we have assume that $A\neq -1$. Now we shall discuss the situation when $A=-1$.\\

The expression for energy density now becomes
\begin{equation}
\rho=\left[3(n+1)Bln(\frac{a}{a_0})\right]^{\frac{1}{(n+1)}},
\end{equation}

which from the first Friedmann equation gives
\begin{equation}
a=a_0 exp\left[b_0\left(t-t_0\right)^{(\frac{1}{1-\alpha})}\right]~,~~b_0=\left(\frac{\sqrt{3}}{2}B(2n+1)\right)^{(\frac{1}{1-\alpha})}.
\end{equation}

From the solutions (5) and (18) we see (fig 1 and 2) that $a\rightarrow 0$ as $t\rightarrow -\infty$, so it is not possible to have emergent scenario with the usual modified Chaplygin gas. However, if we choose $-1<n<-\frac{1}{2}$ then $\alpha>1$ and we have from solution (18) $a\rightarrow a_0$ as $t\rightarrow -\infty$ (see fig 3.). Hence it is possible to have emergent scenario with this revised form of MCG.\\

We shall now discuss the thermodynamics of the emergent scenario with this revised form of  MCG as the cosmic substrum.\\

Assuming the validity of the first law of thermodynamics at the horizon (having area radius $R_h$) we have the Clausius relation:
\begin{equation}
-d E_h=T_h d S_h,
\end{equation}\\

where $T_h$ is the temperature of the horizon and $s_h$ is the entropy of the horizon. In the above, $E_h$ is the amount of energy crossing the horizon during time $dt$ and is given by [23--25]

\begin{figure}
\begin{minipage}{0.4\textwidth}
\includegraphics[width=1.0\textwidth]{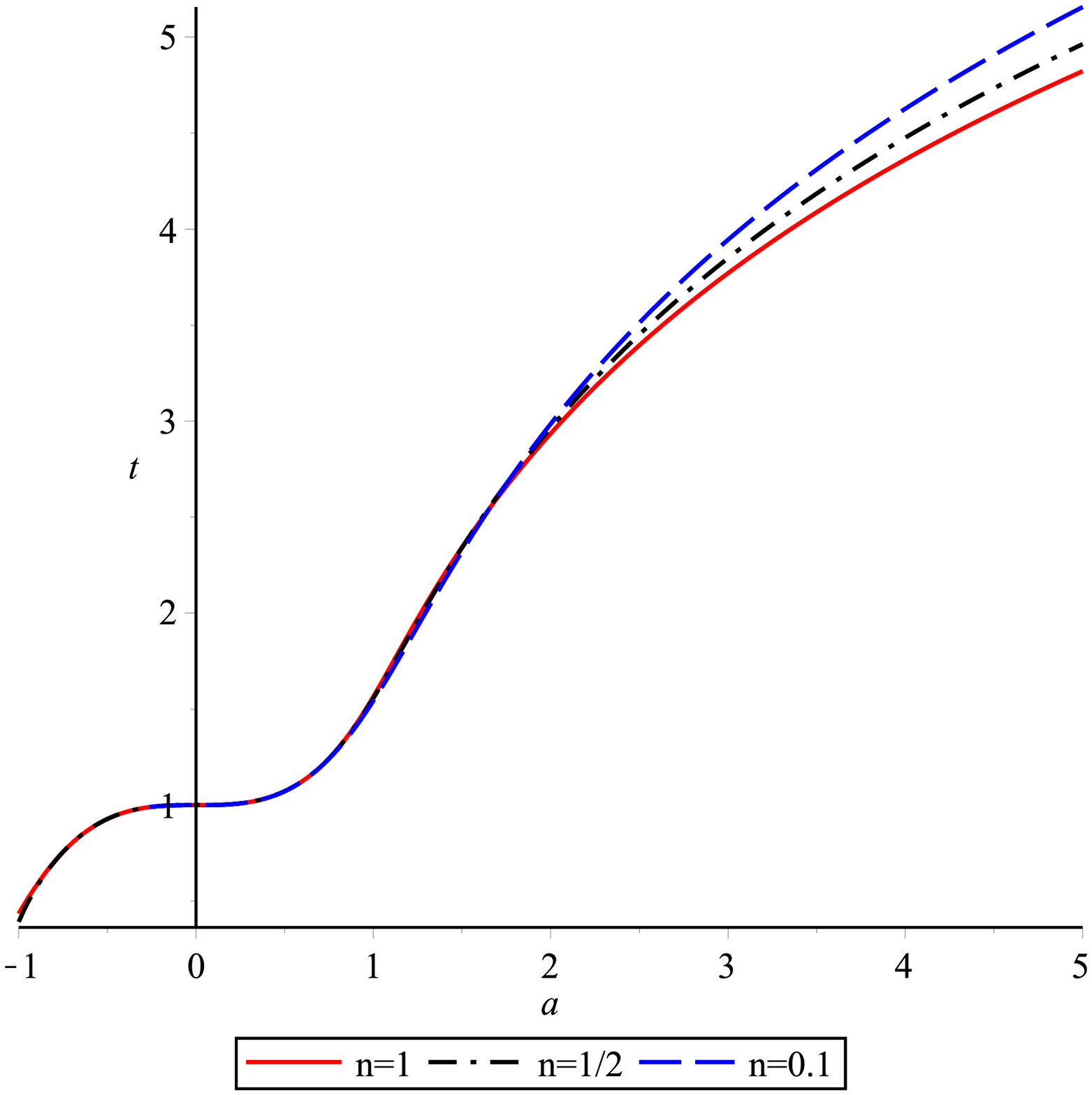}\\
Figure 1: represents the scale factor  $a(t)$ against $t$ for $A\neq-1$.
\end{minipage}
\end{figure}

\begin{figure}
\begin{minipage}{0.4\textwidth}
\includegraphics[width=1.0\textwidth]{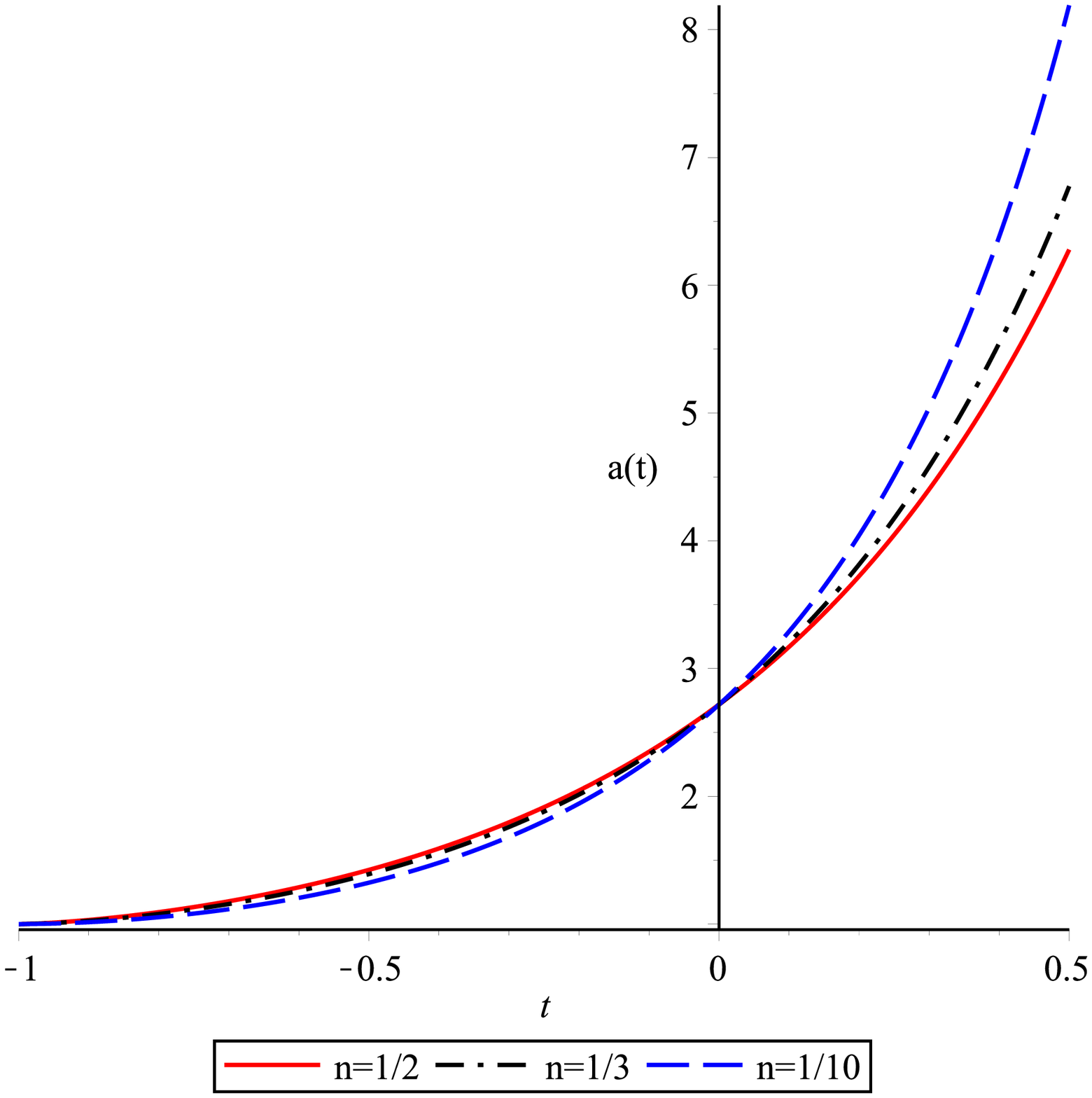}\\
Figure 2: shows the graphical representation of the scale factor  $a(t)$ against $t$ for $A=-1$.
\end{minipage}
\end{figure}

\begin{figure}
\begin{minipage}{0.4\textwidth}
\includegraphics[width=1.0\textwidth]{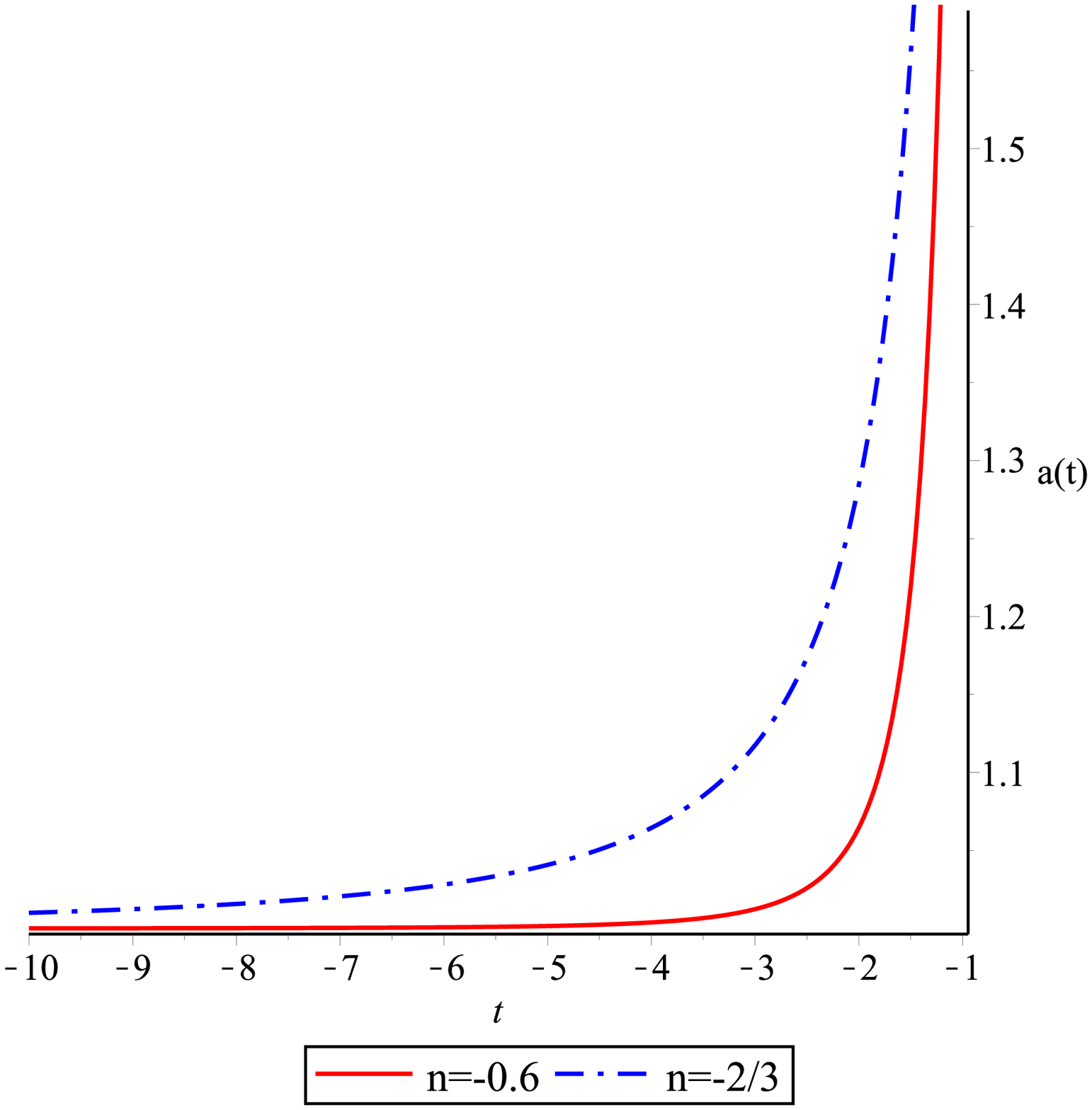}\\
Figure 3: shows the graphical representation of the scale factor  $a(t)$ against $t$ for $A=-1$.
\end{minipage}
\end{figure}

\begin{equation}
-dE_h=4\pi R^3_h H(\rho+p)dt.
\end{equation}

So using equation (20) in (19) we have the rate of change of the horizon entropy as

\begin{equation}
\frac{dS_h}{dt}=\frac{4\pi R^3_h H(\rho+p)}{T_h}.
\end{equation}

To obtain the entropy of the inside fluid we start with the Gibbs equation [25, 26]

\begin{equation}
T_hdS_f=dE_f+p dV,
\end{equation}

where $S_f$ is the entropy of the fluid bounded by the horizon and $E_F$ is the energy of the matter distribution. Here for thermodynamical equilibrium, the temperature of the fluid is taken as that of the horizon i.e, $T_h$.\\

Now using $V=\frac{4\pi R^3_h}{3}$~,~$E_f=\frac{4\pi R^3_h}{3} \rho$ and the Friedmann equations the entropy variation of the fluid is given by

\begin{equation}
\frac{dS_f}{dt}=\frac{4\pi R^2_h}{T_h}(\rho+p)\left(\dot{R_h}-HR_h\right).
\end{equation}

Thus combining (21) and (23) the variation of the total entropy $(S_T)$ is given by

\begin{equation}
\frac{dS_T}{dt}=\frac{d}{dt}(S_h+S_f)=\frac{4\pi R^2_h}{T_h}(\rho+p)\dot{R_h}.
\end{equation}

\underline{{\bf Case-I:}} {\bf Apparent horizon:}\\

The area radius for apparent horizon is given by

$$R_A=\frac{1}{H},$$

so that

$$\dot{R}_A=-\frac{\dot{H}}{H^2}=\frac{4 \pi G(\rho+p)}{H^2}.$$

Hence,

$$\frac{dS_T}{dt}=\frac{(4\pi)^2G(\rho+p)^2}{T_A H^4}>0.$$

Thus generalised second law of thermodynamics (GSLT) is always true at the apparent horizon.\\

\underline{{\bf Case-II:}} {\bf Event horizon:}\\

The area radius for event horizon is given by

$$R_E=a\int^\infty_t \frac{dt}{a}.$$

The above improper integral converges for accelerating phase of the FRW model. Hence in the present scenario it is very much relevant. From the above definition \\

$$\dot{R}_E=HR_E-1,$$

so from (24)

\begin{eqnarray}
\frac{dS_T}{dt}&=&\frac{(4\pi) R_E^2(\rho+p)}{T_E} (HR_E-1),\nonumber\\
&=&\frac{(4\pi) R_E^2 H}{T_E} \left[(A+1)\rho-\frac{B}{\rho^n}\right](R_E-R_A),\nonumber\\
&=&\frac{(4\pi) R_E^2 H}{T_E}\frac{c(1+a)}{a^{3\mu}. \rho^n}(R_E-R_A).\nonumber
\end{eqnarray}

Hence validity of GSLT is possible if $R_E>R_A$ (as $\alpha>1$). In the above temperature is chosen as the hawking temperature on the horizon as [27, 28]

$$T_A=\frac{1}{2\pi R_A}~,~~T_E=\frac{R_E}{2\pi R_A^2}.$$

 In the present work we have examined the cosmology of the emergent  scenario for modified Chaplygin gas as the cosmic fluid. It is found that for both the solutions (with $A\neq -1$ and $A=-1$) the model does not exhibit emergent scenario at early epochs. So one can conclude that it is not possible to have emergent scenario with MCG. However, if $n$ is chosen to be negative  i.e, $-1<n<-\frac{1}{2}$ then $a\rightarrow a_0$ as $t\rightarrow -\infty$ i.e, initial big bang singularity is avoided.\\

Finally, thermodynamical analysis of the emergent scenario has been presented.
\section*{Acknowledgement}
Author SC thanks Inter University Center for Astronomy and Astrophysics (IUCAA), Pune, India for their warm hospitality as a part of the work was done during a visit. Also SC thanks UGC-DRS programme at the Department of Mathematics, Jadavpur University. Author SD thanks Department of Science and Technology (DST), Govt. of India for awarding Inspire research fellowship.


\frenchspacing
\begin{thebibliography}{100}
\bibitem{hr}  E. R. Harrison, {\it Mon.Not.Roy.Astron.Soc} {\bf 137}, 69 (1967).
\bibitem{eg1} George F R Ellis and Roy Maartens, {\it Class.Quant.Grav.} {\bf 21}, 223 (2004).
\bibitem{eg2}  George F.R. Ellis et al., {\it Class.Quant.Grav.} {\bf 21}, 233 (2004).
\bibitem{ms}  S. Mukherjee,  et al. gr-qc/0505103.
\bibitem{ms1} S. Mukherjee,  et al., {\it Class.Quant.Grav.} {\bf 23}, 6927 (2006).
\bibitem{gw}  G. W. Gibbons, {\it Nucl.Phys. B} {\bf 292}, 784 (1988).
\bibitem{gw1} G. W. Gibbons, {\it Ibid} {\bf 310}, 636 (1988).
\bibitem{md}  David J. Mulryne, et al., {\it Phys.Rev. D} {\bf 71}, 123512 (2005).
\bibitem{ab}  A.Banerjee, T.Bandyopadhyay and S.Chakraborty., {\it Gravitation and Cosmology} {\bf 13}, 290 (2007).\\
A.Banerjee, T.Bandyopadhyay and S.Chakraborty., {\it Gen.Relt.Grav.} {\bf 40}, 1603 (2008).
\bibitem{nj}  Nelson J. Nunes, {\it Phys.Rev. D} {\bf 72}, 103510 (2005).
\bibitem{je}  James E. Lidsey and David J. Mulryne, {\it Phys.Rev. D} {\bf 73}, 083508 (2006).
\bibitem{ud}   U.Debnath, {\it Class.Quant.Grav.} {\bf 25}, 205019 (2008).
\bibitem{bp}  B. C. Paul and S.Ghose., {\it Gen.Relt.Grav.} {\bf 42}, 795 (2010).
\bibitem{ud1}  U. Debnath, et al., {\it Int.J.Theor.Phys.} {\bf 50}, 2892 (2011).
\bibitem{sm}   S. Mukerji, et al., {\it Int.J.Theor.Phys.} {\bf 50}, 2708 (2011).
\bibitem{pl}    Pedro Labrana, {\it Phys.Rev. D} {\bf 86}, 083524 (2012).
\bibitem{sc}   S.Chakraborty, {\it Phys. Letts. B} {\bf 81}, 732 (2014).
\bibitem{bp1}  B. C. Paul and A. Majumdar, {\it Class.Quant.Grav.} {\bf 32}, 115001 (2015).
\bibitem{sd}    Sergio del Campo, Ramón Herrera, Diego Pavón, {\it Phys. Letts. B} {\bf 707}, 8 (2012).
 \bibitem{p}  Diego Pavon, et al. arXiv:1212.6863 [gr-qc]
\bibitem{ud2}  U. Debnath, A. Banerjee and S. Chakraborty, {\it Class.Quant.Grav.} {\bf 21}, 5609 (2004).
\bibitem{hb}   H. Benaoum, {\it Preprint hep-th} 10205140 (2002).
\bibitem{rc}   R. G. Cai and S. P. Kim, {\it J. High Energy Physics} {\bf 02}, 050 (2005).
\bibitem{rb}  R. S. Bousso, {\it Phys.Rev. D} {\bf 71}, 064024 (2005).
\bibitem{ns} N. Majumder and S. Chakraborty, {\it Class.Quant.Grav.} {\bf 26}, 195016 (2009).
\bibitem{gd}  G. Izquierdo and D. Pavan, {\it Phys. Letts. B} {\bf 633}, 420 (2006).
\bibitem{sc1}  S. Chakraborty, {\it Phys. Letts. B} {\bf 718}, 276 (2012).
\bibitem{sc2}  S. Saha and S. Chakraborty, {\it Phys. Letts. B} {\bf 717}, 319 (2012).





\end{thebibliography}
\end{document}